\documentclass[12pt]{article}
\usepackage[toc,page]{appendix}
\usepackage{latexsym}
\usepackage{mathrsfs}
\usepackage{amssymb}
\usepackage{amsmath}
\usepackage{amsthm}
\usepackage{amsfonts, epsf,epsfig,color}
\usepackage{graphicx}
\usepackage{tensor}
\usepackage{manfnt}
\usepackage[all]{xy}
\usepackage{titletoc}%
\include{macro}

\newcommand{\C}{\mathbb{C}}

\renewcommand{\P}{\mathbb{P}}

\renewcommand{\P}{\mathbb{P}}

\newcommand{\rd}{\, \mathrm{d}}

\newcommand{\be}{\begin{equation}\label}
\newcommand{\ee}{\end{equation}}
\newcommand{\bea}{\begin{eqnarray}\label}
\newcommand{\eea}{\end{eqnarray}}

\begin{document}

\begin{titlepage}
\rightline{}
\begin{center}
\vskip 2.5cm
{\huge \bf {
Ambitwistor String Theory in the Operator Formalism}}\\
\vskip 2.0cm
{\Large {  R. A. Reid-Edwards  }}
\vskip 0.5cm
{\it {The Milne Centre for Astrophysics \&}}\\
{\it {Department of Physics and Mathematics}}\\
{\it {University of Hull, Cottingham Road}}\\
{\it { Hull, HU6 7RX, U.K.}}\\
{\it {email: r.reid-edwards@hull.ac.uk}}

\vskip 2.5cm
{\bf Abstract}
\end{center}

\vskip 0.5cm

\noindent
\begin{narrower}
After a brief overview of the operator formalism for conventional string theory, an operator formalism for ambitwistor string theory is presented. It is shown how tree level supergravity scattering amplitudes are recovered in this formalism. More general applications of this formalism to loop amplitudes and the construction of an ambitwistor string field theory are briefly discussed.

\end{narrower}

\end{titlepage}

\newpage

\newpage

\section{Introduction}

The operator approach to string theory dates back to the early studies of dual models \cite{DelGiudice:1971fp,Mandelstam:1974fq,Mandelstam:1974hk,Mandelstam:1973jk,Alessandrini:1972ue,Corrigan:1972ux}. The central idea is to encode $N$-point interactions in terms of a vertex $\langle\Sigma_{N,g}|$ which maps $N$ physical states $|\phi\rangle$ into a complex function which is then identified as the scattering amplitude for those states;
$$
\langle V_1(z_1)...V_N(z_N)\rangle=\left\langle \Sigma_{N,g}:(z_1,...,z_N)\right||\phi_1\rangle...|\phi_N\rangle,
$$
where $V_i$ are vertex operators. $N$ and $g$ denote the number of punctures in, and the genus of, the Riemannn surface $\Sigma$. Only the tree level contributions are considered in this paper and so $g=0$ will be assumed throughout. The locations of the punctures $z_i$ will also be supressed. The operator approach played an important role in the discovery that dual models models describe the scattering of strings. After the Polyakov path integral became the staple approach to computing scattering amplitudes in string theory, the operator formalism found new life in the search for a second quantised formulation of string theory that ultimately resulted in string field theory. To set the scene, consider the scattering of $N$ tachyon states $|k\rangle=e^{ik\cdot x}|0\rangle$. Using the standard oscillator expansion for the embedding fields
$$
\partial X^{\mu}(z)=\sum_n\alpha^{\mu}_n\,z^{-n-1},
$$
and associating to each puncture at the point $z_i$ ($i=1,2,..,N$) a Fock space ${\cal H}_i$ with a set of oscillators $\{(\alpha^{(i)}_n)^{\mu}\}$, the vertex can be shown to take the form
\begin{equation}\label{XX}
\langle \Sigma_{0,N}|\sim \langle p_1|...\langle p_N|\;\exp\left(\sum_{i,j=1}^N\sum_{m,n\geq 0}{\cal N}_{mn}(z_i,z_j)\,\alpha^{(i)}_m\cdot\alpha^{(j)}_n\right),
\end{equation}
where $N_{mn}$ are functions of the insertion points $z_i$. The only component in ${\cal N}_{mn}$ of relevance for the $N$ tachyon scattering is\footnote{Assuming $i\neq j$. The $i=j$ term contribution cancels out in the final answer.} ${\cal N}_{00}(z_i,z_j)\sim \ln|z_i-z_j|$ and so using the fact that $\alpha^{(i)}_0|k_j\rangle=\delta_{ij}k_j|k_j\rangle$, it is straightforward to show that
$$
\langle V_1...V_N\rangle=\langle \Sigma_{0,N}||k_1\rangle...|k_N\rangle\sim \prod_{i<j}|z_i-z_j|^{k_i\cdot k_j},
$$
as found from the $X^{\mu}(z)$ contribution to the, more conventional, path integral computation of the amplitude. There are many details that have been skipped over, some of which shall be discussed later on, many more of which are contained in the references.

An important perspective has been provided by the operator formalism for bosonic and supersymmetric string theories and an operator formulation of ambitwistor string theory, a close relative of the type II superstring, will be explored here. In \cite{Mason:2013sva} a string theory describing the embedding of a worldsheet into ambitwistor space was introduced\footnote{See also  \cite{Berkovits:2013xba} for a pure spinor perspective.} and it is thought that, at genus zero, this string theory describes classical type II supergravity in complexified spacetime $\C^{10}$. The key evidence for this conjecture is that ambitwistor string theory reproduces the Cachazo, He and Yuan (CHY) \cite{Cachazo:2014xea,Cachazo:2014nsa,Cachazo:2013iea,Cachazo:2013hca} formulation of tree level scattering amplitudes of type II supergravity. A more direct connection with the supergravity equations of motion has been found by considering the vanishing of the BRST anomaly \cite{Adamo:2014wea}. The status of higher genus amplitudes is still under investigation \cite{Adamo:2013tsa,Geyer:2015jch,Geyer:2015bja,Adamo:2015hoa} and provides part of the motivation for reconsidering this theory in an operator formalism. There are now a number of proposed ambitwistor string theories \cite{Casali:2015vta,Geyer:2014fka,Ohmori:2015sha}, reflecting the myriad application of the scattering equations (\ref{SE}) central to the CHY formalism \cite{Cachazo:2014xea}. Only the original Type II ten-dimensional theory will be considered here; however, the formulation discussed here can be generalised straightforwardly to the other, more exotic, cases.

The primary motivation for this paper is to begin to put in place the tools that will allow the ambitwistor string theory be be studied as a toy model for conventional string theory. This paper presents the first step  in a re-evalution of a number of approaches to studying background independence in string theory that, though initially promising, have proven to be fiendishly difficult due to our as yet poor understanding of string theory in more general curved backgrounds. Future work will consider the study of background independence and the symmetries of supergravity from the perspective of the ambitwistor worldsheet theory. It is  hoped that this study will shed some light on these and other issues in conventional superstring theory.

A secondary motivation is to highlight the differences between conventional string theory and the ambitwistor string. Though similar in many ways, there are stark differences between the theories that become particularly apparent in the operator formulation.

The outline of this paper is as follows: In the rest of this section the original ambitwistor string theory of \cite{Mason:2013sva} and the operator formalism for conventional bosonic string theory are reviewed. In section two an operator formalism for the bosonic matter sector of the ambitwistor string is presented. It is in the bosonic sector that the ambitwistor and conventional superstring theories differ most and this distinction is particularly stark in the operator formalism. To describe scattering amplitudes for conventional Einstein gravity it is necessary to include fermions; this is the subject of section three. Section four rounds off the paper with a discussion of future work, the route to a string field theory description of supergravity, and possible applications to the study of loop amplitudes and background independence.

\subsection{Ambitwistor string theory}

In this section salient features of the ambitwistor string construction of \cite{Mason:2013sva} are reviewed. The starting point is the chiral embedding of a genus zero Riemann surface $\Sigma$ into the cotangent bundle of complexified, ten-dimensional, flat spacetime with coordinates $(X,P)$. After gauge fixing the worldsheet metric\footnote{One might take the worldsheet gravity to be of the `half-twisted' kind suggested in \cite{Mason:2007zv} and discussed further in \cite{ReidEdwards:2012tq}.}, the action of the theory is
$$
S[X,P]=\frac{1}{2\pi}\int_{\Sigma}P_{\mu}\bar{\partial}X^{\mu}+\frac{1}{2}\eta_{\mu\nu}\Psi_r^{\mu}\bar{\partial}\Psi^{\nu}_r+\frac{e}{2}P^2-\chi_rP_{\mu}\Psi^{\mu}_r+b\bar{\partial} c,
$$
where the repeated index $r=1,2$ is summed over and $c$ and $b$ are a standard ghost system resulting from the gauge-fixing of the metric. All fields are chiral. It is assumed from the outset that the theory is in the critical dimension and is conformal\footnote{The vanishing of the central charge as discussed in \cite{Mason:2013sva}.} and the worldsheet metric may be replaced with holomorphic and anti-holomorphic components of the complex structure. Following the constructions \cite{Mason:2007zv,ReidEdwards:2012tq}, only the component of the complex structure that governs the deformation of $\bar{\partial}$ is considered. This component $J$ enters into the above action as $\bar{\partial}=\partial_{\bar{z}}+J\partial_z$. Varying the action with respect to $J$ and then gauge-fixing $J$ to vanish gives the holomorphic stress tensor $T(z)=P_{\mu}\partial X^{\mu}+...$, where the ellipsis denote ghost  and fermion contributions. The stress tensor is expected to play an important role in the construction of an ambitwistor string field theory which will be briefly discussed in section 4.

An additional symmetry, generated by $\delta X^{\mu}=\alpha P^{\mu}$ and $\delta e=\bar{\partial}\alpha$ (all other fields being invariant), reduces the target space to the quotient $P\mathbb{A}$ which is identified with the space of null geodesics, or ambitwistor space \cite{LeBrun,Baston:1987av,LeBrun:1991jh,Isenberg:1978kk}. As described in \cite{Adamo:2013tsa}, the gauge fixing of this residual symmetry is fixed by introducing the gauge fixing fermion
\begin{equation}\label{gauge}
F(e)=e-\sum_{I=1}^{N-3}s_I\mu_I
\end{equation}
where $N$ is the number of punctures, each located at $z_i$, on the genus zero worldsheet $\Sigma$ and $\mu_I$ is a basis for $H^{0,1}(\Sigma, T_{\Sigma}(-z_1-...-z_N))$. Introducing ghosts and Nakanishi-Lautrup fields in the standard way gives rise to a standard ghost system with fields $\tilde{b}$ and $\tilde{c}$ of conformal weight $+2$ and $-1$ respectively and  an insertion of
$$
\prod_{I=1}^{N-3}\bar{\delta}\left(\int_{\Sigma}\mu_I\,P^2\right)\;\int_{\Sigma}\tilde{b}\,\mu_I,
$$
into the path integral  \cite{Adamo:2013tsa}. The second integral term is the counterpart of the standard $b$ ghost insertion in conventional string theory and absorbs $N-3$ $\tilde{c}$ ghost insertions on the vertex operators. The first term plays a role that has no direct counterpart in conventional string theory. The expression for an $N$-point amplitude is then
$$
{\cal M}(1,...,N)= \int_{\Gamma\subset{\cal M}_{0,N}}\left\langle\prod_{I=1}^{N-3} \bar{\delta}\left(\int_{\Sigma}\mu_I\,P^2\right)\;\int_{\Sigma}\tilde{b}\,\mu_I\;\int_{\Sigma}b\,\mu_I\; {\prod_{i=1}^N}c\tilde{c}\,V_i(z_i)\right\rangle,
$$
where the vertex operators in the NS sector are given by
$$
V(z)= e^{ik\cdot X(z)}\prod_{r=1,2}\left(\epsilon_r^{\mu}P_{\mu}+k_{\mu}\Psi_r^{\mu}\;\epsilon_{r\nu}\Psi^{\nu}_r\right).
$$
and $\Gamma$ is a $N-3$ dimensional cycle in the $2N-6$ dimensional moduli space ${\cal M}_{0,N}$ of the $N$-punctured sphere. As argued in \cite{Adamo:2013tsa}, the $N-3$ $\tilde{c}$ ghost terms combine with the $N-3$ integrations over the location of the punctures, the only moduli on a $N$-punctured sphere, to give the integrated vertex operators
$$
\int_{\Sigma}\bar{\delta}(k^{\mu}P_{\mu})V(z).
$$
The amplitude then becomes
$$
{\cal M}(1,...,N)= \left\langle\prod_{i=1}^3c\tilde{c}V_i(z_i)\prod_{i=4}^N\int_{\Sigma}\rd^2z_i\,\bar{\delta}(k_i\cdot P)V_i(z_i)\right\rangle.
$$
Evaluating the remaining ghost contributions gives two copies of the volume element\footnote{The $c$ ghosts give a holomorphic contribution to Vol SL$(2;\C)$. The anti-holomorphic contribution comes from the $\bar{c}$ ghosts of the half-twisted worldsheet gravity which is included implicitly.} $\rd V_{ijk}$, given in (\ref{Vol}). Performing the, thus far suppressed, fermionic integrals gives the NS sector scattering amplitudes \cite{Cachazo:2014xea,Cachazo:2014nsa,Cachazo:2013iea,Cachazo:2013hca}
\begin{equation}\label{CHY}
{\cal M}(1,...,N)=\delta^D\left(\sum k_i\right)\int_{\Gamma\subset{\cal M}_{0,N}}\frac{1}{\text{Vol SL}(2;\C)}\text{Pf}'(M_1)\text{Pf}'(M_2)\prod_i'\bar{\delta}\left(k_i\cdot P(z_i)\right),
\end{equation}
where $P(z_i)$ is given, after integrating out the $X$ non-zero modes,  by
\begin{equation}\label{P1}
P_{\mu}(z)=\sum_{j=1}^N\frac{k^j_{\mu}}{z-z_j}.
\end{equation}
The delta functions $\bar{\delta}\left(k_i\cdot P(z_i)\right)$ impose the scattering equations
\begin{equation}\label{SE}
\sum_{j\neq i}\frac{k_i\cdot k_j}{z_i-z_j}=0,
\end{equation}
which originally appeared in \cite{Fairlie:1972zz,Gross:1987ar}. The objects Pf$'M_r$, $r=1,2$, encode the polarisation data and their relationship to the Pfaffian of the matrix $M_r$ and the definition of $M_r$ are given in \cite{Mason:2013sva}.

\subsection{The operator formalism}

In this section certain aspects of the operator formalism as applied to the conventional bosonic string are reviewed. In the next section similar constructions shall be given for the ambitwistor string reviewed above. The central tenet of the operator formalism is to recast the scattering amplitude, written in terms of a path integral with vertex operator insertions $V(z_i)$ at punctures $z_i$, as an inner product of $N$ external between states $|\Phi\rangle$ and the principle object; the vertex $\langle \Sigma_{0,N}|$, so that
$$
\langle V_1(z_1)...V_N(z_N)\rangle=\langle \Sigma_{0,N}||\Phi_1\rangle\otimes...\otimes |\Phi_N\rangle.
$$
The state $|\Phi_i\rangle$ is related to the vertex operator $V_i(z_i)$ by
$$
|\Phi_i\rangle=\lim_{t_i\rightarrow 0}:V_i(t_i)|0\rangle,
$$ 
where $t_i$ is a local coordinate system chosen around each puncture such that the location of the $i$'th puncture is $t_i=0$ and so the oscillator description of the vertex operator takes the standard form when written in these $t_i$ coordinates. Note that the information on the location of the puncture is carried by the vertex operator in the path integral whereas it is contained in the vertex $\langle \Sigma_{N}|$ in the operator formalism. Formally the vertex $\langle \Sigma_{N}|$ is a map $V:\otimes_{i=1}^N {\cal H}_i\rightarrow \C$ from an $N$-fold product of Fock spaces, one based at each puncture. There are many ways of deriving the the vertex\footnote{An incomplete sample is \cite{DiVecchia:1986mb, DiVecchia:1986uu, Neveu:1987xb, LeClair:1987gj,Neveu:1986um, Neveu:1985gx}. The relationships between some of these approaches is explored in \cite{DiVecchia:1987jb,Ohrndorf:1989bz,Ohrndorf:1988ui}.} $\left\langle  \Sigma_{0,N}\right|$; the one followed here admits a relatively simple generalisation to higher genus Riemann surfaces and so is of particular interest \cite{DiVecchia:1987jb,Vafa:1987es, AlvarezGaume:1988bg, AlvarezGaume:1989qq}. The general idea is to find conserved charges $Q_n$, where $n\geq 0$, which preserve the vertex
$$
\langle \Sigma_N\hspace{-.1cm}:\hspace{-.1cm}X|\,Q_n=0.
$$
The notation $\langle \Sigma_N\hspace{-.1cm}:\hspace{-.1cm}X|$ is intended to stress the fact that the vertex is for the contribution given by the scalar field $X(z)$ and not the full theory. Knowing the form of the $Q_n$ then specifies the vertex up to an overall constant. The specification of $Q_n$ is perhaps best seen from the path integral description for the associated transition amplitude, viewed as a CFT defined on a Riemann surface with $N$ discs $\mathscr{D}$ removed. For a Riemann surface with a single disc removed, the path integral over this surface is a wave functional $\Psi[X_{\text{cl}}]=\langle\Sigma_{1}|X\rangle$
$$
\Psi[X_{\text{cl}}]=\int_{X|_{\partial\Sigma}}{\cal D}X\; e^{-S[X]},
$$
 where $X_{\text{cl}}=X|_{\partial\Sigma}$ is the classical state of the string on the boundary of the removed disc. Taking the double of the surface\footnote{The double of a Riemann surface is in this case the Riemann surface with the disc filled in.} and denoting the point on which the disc is centred as $p$, then a conserved charge may be written in terms of the current $J_n=f_n\partial X$ where $f_n$ is holomorphic on $\Sigma\backslash\{\mathscr{D}\}$ and has a pole of degree $n$ at $p$. The effect of the transformation induced by $Q_n$ is the symmetry $\Psi[X_{\text{cl}}+f_n]=\Psi[X_{\text{cl}}]$, where the function $f_n$ is given by
$$
f_n(t)=t^{-n}+\sum_{m\geq 0}{\cal N}_{nm}\,t^m,
$$
for some coefficients ${\cal N}_{nm}$, where $t$ is a local coordinate that vanishes at $p$. The key point is that $f_n$ has a pole of order $n$ at the puncture $t(p)=0$ and is holomorphic outside a disc surrounding $p$. 

For a Riemann surface with several discs removed, it is useful to generalise the function $f_n$ and introduce a local coordinate $t_i$ which vanishes at the centre of the $i$'th disc (the location of the  $i$'th puncture). The standard operator approach is to take a local coordinate system $t_i$ centred on the $i$'th puncture and relate it to a coordinate $z$ on the complex plane by a conformal mp $z=h_i(t_i)$. The location of the $i$'th puncture is then given by $z_i=h_i(0)$. Again, the key point is that the harmonic function $f^i_n$ only has a pole (of order $n$) at $t_i$ and is regular at other points. Introducing the inverse map $\zeta_i(z)=t_i$, $f_n$ may be written as
\begin{equation}\label{f}
f^i_n(z)=\delta^{ij}\zeta_j^{-n}+\sum_j\sum_{m=1}^{\infty}{\cal N}_{nm}(z_i,z_j)\,\zeta_j^{m-1}.
\end{equation}
The functions ${\cal N}_{nm}(z_i,z_j)$ may be calculated straightforwardly and are given in the Appendix. It is useful to think of\footnote{The Riemann surface is now thought of as having $N$ punctures with a classical state $X_{c_i}$ on the boundary of the disc surrounding the $i$'th puncture.} $\Psi[\{X_{\text{cl}}^{(i)}\}]$ as the inner product of the $N$ boundary states $|X_{\text{cl}}^{(i)}\rangle$ and the vertex $\langle \Sigma_N\hspace{-.1cm}:\hspace{-.1cm}X|$ and to then recast the invariance of $\Psi[\{X_{\text{cl}}^{(i)}\}]$ in terms of a set of operator equations which $\langle \Sigma_N\hspace{-.1cm}:\hspace{-.1cm}X|$ must satisfy. Around each puncture, the fields have a standard oscillator expansion in the $t$ coordinate 
\begin{equation}\label{x}
\partial X^{\mu}=\sum_n\alpha_n^{\mu}\,t^{-n-1},
\end{equation}
and it is useful to introduce a Fock space of oscillators $\alpha^{(i)}_n$ around the $i$'th puncture so that $X=\sum_iX^{(i)}$ where\footnote{There are some subtleties in some of the terms included in the $+...$. A discussion of this and related issues may be found in \cite{DiVecchia:1986mb}.} $X^{(i)}(z)=X^{(i)}(h_i(t_i))+...= p_{(i)}\ln(z-z_i)+...$. Substituting in (\ref{f}) and (\ref{x}), the charges $Q_n=\oint\rd z\partial X\;f_n$ may be written as $\alpha^{(i)}_{-n}+\sum_jN_{nm}(z_i,z_j)\alpha_m^{(j)}$. The requirement that the vertex $\langle \Sigma_N\hspace{-.1cm}:\hspace{-.1cm}X|$ is invariant under the transformation generated by $Q_n$ requires that
\begin{equation}\label{Q}
\langle \Sigma_N\hspace{-.1cm}:\hspace{-.1cm}X|\,Q_n=\langle \Sigma_N\hspace{-.1cm}:\hspace{-.1cm}X|\left(\alpha_{-n}^{(i)}+\sum_{m,j}{\cal N}_{nm}(z_i,z_j)\,\alpha_m^{(j)}\right)=0.
\end{equation}
It is useful to make the replacement
$$
\alpha^{(i)}_{-n}\rightarrow\frac{\partial}{\partial y^i_n},	\qquad	\alpha^{(i)}_n\rightarrow ny^i_n,	\qquad	n\geq 0,
$$
which reproduces the commutator $[\alpha^{(i)}_m,\alpha^{(j)}_n]=n\delta^{ij}\delta_{m+n,0}$. Care must be taken when dealing with the zero modes \cite{DiVecchia:1986mb,Neveu:1987xb}, but one can eventually show that the vertex that satisfies (\ref{Q}) is
\begin{equation}\label{Vbos}
\langle \Sigma_N\hspace{-.1cm}:\hspace{-.1cm}X|=C\int\prod_i\rd^Dp_i\;\delta\left(\sum p_i\right) \;\langle p_1|...\langle p_N|\; \exp\Bigg(\frac{1}{2}\sum_{i,j}^N\sum_{m,n=0}^{\infty} {\cal N}_{nm}(z_i,z_j)\,\alpha^{(i)}_m\cdot \alpha^{(j)}_m\Bigg),
\end{equation}
where $C$ is a constant that will be neglected from now on.

To complete the description of the bosonic string ghost contributions must be included. Similar arguments to those above may be used to determine the contribution of the ghosts to the vertex to give $\langle \Sigma_N\hspace{-.1cm}:\hspace{-.1cm}X,b,c|$ which, since the contributions of all fields in the theory are now included, may be unambiguously written as $\langle \Sigma_N|$. The vertex is given by
\begin{eqnarray}\label{V}
\langle \Sigma_N|&=&  \int\prod_{i=1}^N\rd p^i\; \delta^D\left(\sum p^i\right)\;\left\langle p_1,3,\bar{3}\right|...\left\langle p_N,3,\bar{3}\right| \;\prod_{r=-1}^{+1}{\cal Z}_r\;\prod_{s=-1}^{+1}\bar{\cal Z}_s\nonumber\\
 &\times&\exp\left(\frac{1}{2}\sum_{i,j=1}^N\sum_{n,m=0}^{\infty} {\cal K}_{nm}(z_i,z_j)\,\alpha_n^{(i)}\cdot \alpha_m^{(j)}+ \sum_{i,j=1}^N\sum_{\substack{n=2\\ m=-1}}^{\infty} {\cal K}_{nm}(z_i,z_j)\,c_n^{(i)}\, b_m^{(j)}+...\right)
\end{eqnarray}
where the coefficients ${\cal N}_{nm}^{ij}$, ${\cal K}^{ij}_{nm}$ and ${\cal M}_{nm}^i$ are functions of the $z_i$ and are given in Appendix A. The $+...$ in (\ref{V}) denotes corresponding anti-holomorphic contributions. The states $\left\langle 3\right|=\left\langle 0\right|c_{-1}c_0c_1$ are normalised such that $\left\langle 3\right| 0\rangle=1$ and similarly for $\left\langle \bar{3}\right|=\left\langle 0\right|\bar{c}_{-1}\bar{c}_0\bar{c}_1$. The expression for the vertex includes
\begin{equation}\label{H}
{\cal Z}_r=\sum_{i=1}^n\sum_{m=-1}^{\infty}{\cal M}_{r m}(z_i)\,b_m^i.
\end{equation}
The ${\cal Z}_r$ are related to the SL$(2;\C)$ invariance of the theory. In order to compute scattering amplitudes we must introduce on shell states $|\phi\rangle$, defined by the insertion of the corresponding vertex operator $V_{\phi}(t)$ at the origin of a local coordinate system
$$
| \phi\rangle=\lim_{t\rightarrow 0}:c\bar{c}V_{\phi}(t)| 0\rangle.
$$
Of particular interest here are the massless states given by
$$
| \epsilon,k\rangle=\lim_{t\rightarrow 0}:c(t)\bar{c}(t) \epsilon_{\mu\nu}\partial X^{\mu}(t) \bar{\partial}X^{\nu}(t)e^{ik\cdot X(t)}|0\rangle= c_1\bar{c}_1 \epsilon_{\mu\nu} \alpha_{-1}^{\mu} \bar{\alpha}_{-1}^{\nu} e^{ik\cdot x}| 0\rangle.
$$
To calculate scattering amplitudes the moduli - the locations of $N-3$ punctures - must be integrated over, giving
$$
\langle V_{N}|=  \int_{\Gamma\subset{\cal M}_{0,N}}\rd^{2N-6}\mathfrak{m}  \left\langle \Sigma_{N}\right| \prod_{I=1}^{N-3} b_I\,\bar{b}_I,
$$
where
\begin{equation}\label{b}
b_I=\int_{\Sigma}\rd^2 z \,\mu_I\,b(z),
\end{equation}
where $\mu_I$ is a basis of beltrami differentials\footnote{In terms of a local coordinate system $t_i$ near the $i$'th puncture $b(z)$ may be written as
$$
b(z)=(h'_i(t_i))^{-2}\sum_nb^i_nt_i^{-n-2}.
$$} and $\Gamma$ is a suitable $N-3$ dimensional holomorphic cycle. The expression for the amplitudes becomes
$$
{\cal M}(1,...,N)= g^{N-2} \int_{\Gamma\subset{\cal M}_{0,N}}\rd^{2N-6}\mathfrak{m}  \left\langle \Sigma_N\right| \prod_{I=1}^{N-3} b_I\,\bar{b}_I\,| \phi_1\rangle...| \phi_N\rangle,
$$
with $\langle \Sigma_{N}|$ given by (\ref{V}). Though especially useful for higher genus, the conserved charge approach outlined above is not the only way to find the vertex and it will be useful to employ other techniques in the case of the ambitwistor string at tree level.

\section{An operator formalism for ambitwistor string theory}

The aim of this section is to construct a vertex $\langle V_N|$ that reproduces the bosonic part of the ambitwistor $N$-point scattering amplitude \cite{Mason:2013sva}
$$
\langle\Sigma_N||\Phi_1\rangle\otimes...\otimes |\Phi_N\rangle=\left\langle \prod_{i=1}^3c_i\tilde{c}_iV_i \; \prod_{j=4}^N\int_{\Sigma}\bar{\delta}(k_j\cdot P(z_j))V_j\right\rangle.
$$
It will be simplest to decompose $\langle\Sigma_N|$ into a product of matter and ghost sectors
$$
\langle\Sigma_N|=\langle\Sigma_N\hspace{-.1cm}:\hspace{-.1cm}X,P|\otimes\langle\Sigma_N\hspace{-.1cm}:\hspace{-.1cm}b,c|\otimes\langle\Sigma_N\hspace{-.1cm}:\hspace{-.1cm}\tilde{b},\tilde{c}|
$$
and to treat each individually.

\subsection{The matter sector}

The starting point is the gauge-fixed action, with $N$ tachyonic vertex operator insertions
\begin{equation}\label{T}
S[X,P]=\int_{\Sigma}P\cdot \bar{\partial}X+\int_{\Sigma}i\sum_{i=1}^Nk_i\cdot X(z)\delta^2(z-z_i).
\end{equation}
The vertex for general states will be constructed and, as a limit of the general case, the truncated vertex for the scattering of massless states on-shell will be derived. The term \emph{truncated vertex} will be used to mean refer to a vertex for which all operators that play no non-trivial role are omitted\footnote{An operator will be taken to play only a trivial role if it is annihilated by the external states.}. Decomposing $X^{\mu}$ into a zero mode part $x^{\mu}$ and a non-zero mode $\breve{X}^{\mu}$, the functional $\Psi[X,P]$ may be expressed as
$$
\Psi[X,P]=\delta^D\left(\sum k_i\right)\Psi[\breve{X},P],
$$
where the momentum-conserving delta function comes from the $x^{\mu}$ integral as usual. The $\breve{X}$ integral simply enforces the condition that $P(z)$ takes the classical value $P_{\text{cl}}(z)$ given by (\ref{P1}). Using the fact that $\langle p|k\rangle=\delta^D(p-k)$, the conservation of momentum may be incorporated into the vertex straightforwardly and one might be tempted to propose
\begin{equation}\label{O}
\langle\Sigma_N\hspace{-.1cm}:\hspace{-.1cm}X,P|=\int{\cal D}P\int\prod_i\rd^Dp_i\;\delta\left(\sum p_i\right)\,\langle p_1|...\langle p_N|\;\delta[P(z_i)-P_{\text{cl}}(z_i)].
\end{equation}
The imposition of the condition $P(z)=P_{\text{cl}}(z)$ by integrating over a delta-function is not particularly natural from the operator perspective if one is to ultimately consider the momenta as operators. If the states of interest were simply tachyon-like $|k\rangle$, the above expression might suffice for most purposes; however, the on-shell massless states $P_{\mu}P_{\nu}|k\rangle$ directly involve the momenta and so more work is needed to bring (\ref{O}) to a more suitable form.

\subsubsection{The vertex for general sources}

The basic idea is that the amplitude $A_N(\Phi_1,...,\Phi_N)$ corresponding to a worldsheet with $N$ states $\Phi_i$ inserted at the $i$'th puncture may be calculated by $\langle \Sigma_N||\Phi_1\rangle...|\Phi_N\rangle$. The problem is that we do not know what $\langle \Sigma_N|$ is. Following \cite{DiVecchia:1986mb}, where similar arguments were put forward in the context of the conventional bosonic string, the vertex may be determined from the action evaluated on the classical fields. This may be seen by introducing a basis of coherent states $\{|X,P\rangle\}$. The basic idea is that the amplitude for the scattering of $N$ coherent states  $A_N(X_{\text{cl}}^1,...,X^N_{\text{cl}},P_{\text{cl}}^1,...,P^N_{\text{cl}})$ has the same functional form as the vertex with the eigenvalues $X$ taking the place of the operators since, for coherent states, $\widehat{X}|X\rangle=X|X\rangle$ and so we define the function ${\cal F}_N=\langle\Sigma_N\hspace{-.1cm}:\hspace{-.1cm}X,P||X_1,P_1\rangle...|X_N,P_N\rangle$. Since this function is just the amplitude for the scattering of $N$ coherent states, it may equally well be calculated using the path integral approach
$$
{\cal F}_N=\int{\cal D}X{\cal D}P\;e^{S[X,P]}\prod_{i=1}^NV_i,
$$
The question is, what are the vertex operators? The answer is a straightforward generalisation of the simple tachyon scattering problem above (\ref{T}). There, the tachyon vertex operators were incorporated by including a source term $J\cdot X$ where, for the tachyons $J(z)=\sum_ik_i\delta^2(z-z_i)$. The form of the current could have been read off directly from the right hand side of the $X(z)$ equations of motion $\bar{\partial}P_{\text{cl}}=J$, and the source term in the action may be expressed as $X\cdot\bar{\partial} P_{\text{cl}}$, where $P_{\text{cl}}$ is interpreted as the classical field configuration at the puncture. The path integral above is then given by
$$
{\cal F}_N= Z[J,K]=\int{\cal D}X{\cal D}P\;e^{S[X,P]},
$$
where the generating function $Z[J,K]$ has sources $J_{\mu}$ and $K^{\mu}$, sources which are related to the classical field configurations of the coherent states. The gauge-fixed action, with these sources included, is given by
\begin{equation}\label{JK}
S[X,P]=\int_{\Sigma}P\cdot \bar{\partial}X-K\cdot P+J\cdot X,
\end{equation}
where the equations of motion satisfied by the classical fields are $\bar{\partial}P_{\text{cl}}=J$ and $\bar{\partial}X_{\text{cl}}=K,$ and so we may replace $J$ and $K$ by $\bar{\partial}P_{\text{cl}}$ and $\bar{\partial}X_{\text{cl}}$ respectively in (\ref{JK}). Separating out the zero modes $x$ and then integrating out the non-zero modes $\breve{X}$ in the generating functional gives
\begin{equation}\label{New}
Z[J,K]\sim \delta^D\Big(\oint_{\partial \Sigma}\rd zP_{\text{cl}}\Big)\int{\cal D}P\;\delta[P(z)-P_{\text{cl}}(z)]\;e^{-\int_{\Sigma}\rd^2z K(z)\cdot P(z)}
\end{equation}
where the fact that $\bar{\partial}P_{\text{cl}}=J$ has been used. Using the delta functional to do the $P$ integral and replacing the source $K$ with $\bar{\partial}X_{\text{cl}}$ gives
\begin{equation}\label{Me}
Z[J,K]\sim \delta^D\Big(\oint_{\partial \Sigma}\rd zP_{\text{cl}}\Big)\;e^{-S[X_{\text{cl}},P_{\text{cl}}]}
\end{equation}
where
$$
S[X_{\text{cl}},P_{\text{cl}}]=\int_{\Sigma}P_{\text{cl}}\cdot \bar{\partial}X_{\text{cl}},
$$
The $x$ zero modes give rise to a delta function which, as will be discussed below, enforces overall momentum conservation.

Introducing local coordinates $t_i$ around each puncture as described in the introduction, the classical fields may be written as mode expansions in these local coordinates\footnote{It is important not to confuse the $\alpha$ and $\tilde{\alpha}$ appearing here with the operators in the bosonic sector of conventional string theory. In particular, there is no analogue of the anti-holomorphic sector here. Subtleties relating to the $+...$ terms are discussed in \cite{DiVecchia:1986mb}.} 
\begin{equation}\label{Com}
P^{(i)}_{\text{cl}}(z)=(h'_i(t_i))^{-1}\sum_n\alpha_n^{(i)}t^{-n-1}+...,	\qquad	X^{(i)}_{\text{cl}}(z)=\sum_n\tilde{\alpha}_n^{(i)}t^{-n}+...
\end{equation}
The support of the delta-function may then be written as
$$
\sum_j\oint_{z_j}\rd zP_{\text{cl}}(z)=\sum_{i,j}\sum_n\alpha_n^{(i)}\oint_{z_j}\rd z\,(h'_i(t_i))^{-1}\,t^{-n-1}
$$
this in only non-trivial of $i=j$ and, recalling that $\rd z=h_i'(t_i)\rd t_i$, the contour integral may be simply done to yield
$$
\oint_{\partial \Sigma}\rd zP_{\text{cl}}(z)=\sum_{i=1}^Nk_i
$$
where the operator $\alpha_0^{(i)}$ has been identified with the zero mode $k_i$. The delta function therefore gives rise to the expected conservation of momentum. After some manipulation, it may be shown that
$$
S[X_{\text{cl}},P_{\text{cl}}]=\sum_{i,j}\oint_{z_i}\rd z\oint_{z_j}\rd w \,P_{\text{cl}}(z)S(z,w)X_{\text{cl}}(w),
$$
where $S(z,w)$ is the Green's function for $\bar{\partial}$ and $X_{\text{cl}}$ and $P_{\text{cl}}$ have the oscillator expansions (\ref{Com}). Putting this all together one finds that the vertex is
\begin{equation}\label{SFT}
\langle\Sigma_N\hspace{-.1cm}:\hspace{-.1cm}X,P|=C\int\prod_i\rd^Dp_i\;\delta\left(\sum p_i\right) \;\langle p_1|...\langle p_N|\; \exp\Bigg(\sum_{i,j}^N\sum_{m,n=0}^{\infty} {\cal S}_{nm}(z_i,z_j)\,\tilde{\alpha}^{(i)}_m\cdot \alpha^{(j)}_n\Bigg),
\end{equation}
where $P_{\text{cl}}$ and $X_{\text{cl}}$ have oscillator modes $\alpha_n$ and $\tilde{\alpha}_n$ respectively (recall that, in this gauge $X$ and $P$ are independent). $C$ is a constant that will be neglected from now on. The $X$ and $P$ fields are conjugate and so the oscillators satisfy the commutation relations
$$
[\tilde{\alpha}^{(i)\mu}_m,\alpha^{(j)}_{n\nu}]=\delta_{ij}\delta^{\mu}_{\nu}\delta_{m+n,0}.
$$
Defining a map $h_i$ from local coordinates $t_i$ to the coordinate $z$, where $z_i=h_i(0)$ as described above, the vertex function $S_{mn}(z_i,z_j)$ in (\ref{SFT}) is given by
$$
S_{mn}(z_i,z_j)=\oint\frac{\rd t_i}{2\pi i}\oint\frac{\rd t_j}{2\pi i}\;h'_j(t_j)\,t_i^{-m}\,t_j^{-n-1}\,\frac{1}{h_i(t_i)-h_j(t_j)}.
$$
Notice that $S_{0n}(z_i,z_j)=0$ for all $n\geq 0$ and also that $S_{mn}(z_i,z_j)\neq S_{nm}(z_i,z_j)$.

\subsubsection{On-shell scattering vertex}

It is useful to see how the above construction imposes the delta function condition on the momenta in the path integral formalism.The situation simplifies greatly if the sources are restricted to on shell asymptotic states of the form $|k_i,\varepsilon_i\rangle=\alpha_{-1}^{\mu}\alpha_{-1}^{\nu}e^{ik\cdot x}|0\rangle$. Setting $J_{\mu}(z)=i\sum_ik_i\delta^2(z-z_i)$ and localising the momentum source to the punctures by writing $K^{\mu}(z)=\sum_i\zeta^{\mu}_i\delta^2(z-z_i)$, for some set of dummy variables $\zeta^{\mu}_i$, the expression for the vertex (\ref{New}) becomes
$$
{\cal F}_N= \delta^D\Big(\sum_ik_i\Big)\int{\cal D}P\;\delta[P(z)-P_{\text{cl}}(z)]\;e^{-\sum_i\zeta_i\cdot P(z_i)}
$$
The dummy $\zeta_i$ variables allows for differentiation with respect to $\zeta_i$ to bring down powers of $P(z_i)$ as necessary and subsequently set $\zeta_i$ to zero\footnote{This just incorporates the standard path integral technique of writing the vertex operators in the exponential form
$$
 V(z)=\epsilon^{\mu\nu}P_{\mu}(z)P_{\nu}(z)e^{ik\cdot X(z)}=\left[\epsilon^{\mu\nu}\frac{\partial^2}{\partial\zeta^{\mu}\partial\zeta^{\nu}}\exp\left(ik\cdot X(z)+\zeta\cdot P(z)\right)\right]_{\zeta=0}.
$$}. Such terms will be naturally incorporated into the external states and shall not be included in the definition of the vertex and so the momentum source $K$ is set to zero. The classical states are then given by the equations of motion
$$
\bar{\partial}X_{\text{cl}}^{\mu}(z)=0,	\qquad	\bar{\partial}P_{\text{cl}\;\mu}(z)=\sum_{i=1}^Nk_{\mu}^i\delta^2(z-z_i).
$$
There are no globally defined holomorphic vector fields on the worldsheet and so
\begin{equation}\label{P}
P_{\text{cl}}(z)=\sum_{i=1}^N\frac{k^i}{z-z_i}.
\end{equation}
Integrating out the zero mode $x^{\mu}$ gives the required momentum-conserving delta function. After setting the $\zeta_i$ to zero, the $X^{\mu}$ equation of motion is $\bar{\partial}X^{\mu}=0$. Liouville's theorem tells us that $X^{\mu}$ is a constant as it is holomorphic everywhere on the compact Riemann surface $\C\P^1$ and so $X_{\text{cl}}^{\mu}(z)=x^{\mu}$. Since $X^{\mu}_{\text{cl}}=x^{\mu}$, then $S[X_{\text{cl}},P_{\text{cl}}]=0$ and therefore there is no other content to the vertex $\langle\Sigma_N\hspace{-.1cm}:\hspace{-.1cm}X,P|$ other than that from the zero modes which give the momentum conservation, the constraint requiring that $P(z)$ takes the classical value (\ref{P}), and the second term in (\ref{SS}), which is related to SL($2;\C$) invariance of the vacuum. A reasonable first step towards understanding the matter contribution to the vertex was the proposal
$$
\int{\cal D}P \int\prod_i\rd^Dp_i\;\delta\left(\sum p_i\right)\,\langle p_1|\otimes...\otimes\langle p_N|\;\delta[P(z_i)-P_{\text{cl}}(z_i)].
$$
As discussed above, an operator description of this delta functional that forces the momenta to take the classical value (\ref{P}) would be desirable. A straightforward application of the Baker-Campbell-Hausdorf formula\footnote{In the adjoint form
$$
e^XY=\left(Y+[X,Y]+\frac{1}{2}[X,[X,Y]]+...\right)e^X.
$$
The higher terms denoted by the ellipsis do not contribute to the calculation. Indeed, only the $[X,[X,Y]]$ term plays a role.} yields
$$
\langle p_i|\exp\left(P_{\text{cl}}(z_i)\cdot\tilde{\alpha}_1^{(i)}\right)\,\varepsilon_{\mu\nu} \alpha_{-1}^{(i)\mu} \alpha_{-1}^{(i)\nu}|k_i\rangle=\varepsilon_{\mu\nu} P_{\text{cl}}^{\mu}(z_i)P_{\text{cl}}^{\nu}(z_i)\,\delta^D(p_i-k_i).
$$
$P_{\text{cl}}$ takes the value (\ref{P}). The matter contribution to the vertex is then
\begin{equation}\label{X}
\langle\Sigma_N\hspace{-.1cm}:\hspace{-.1cm}X,P|= \int\prod_i\rd^Dp_i\;\delta\left(\sum p_i\right)\,\langle p_1|\otimes...\otimes\langle p_N|\; \exp\left(\sum_{i\neq j}S(z_i,z_j)\,\alpha_0^{(j)}\cdot\tilde{\alpha}_1^{(i)}\right),
\end{equation}
where the Green's function of $\bar{\partial}$ is written as $S(z_i,z_j)=(z_i-z_j)^{-1}$ so that (\ref{P}) becomes $P_{\text{cl}}(z_i)=\sum_{j\neq i}k_iS(z_i,z_j)$. The massless external states at the punctures are
$$
|k_i,\varepsilon_i\rangle=\varepsilon^{\mu\nu}\alpha_{-1\mu}\alpha_{-1\nu}e^{ik\cdot x}|0\rangle,
$$
and one can show that the matter contribution to the vertex (\ref{X}) satisfies
$$
\langle\Sigma_N\hspace{-.1cm}:\hspace{-.1cm}X,P||k_1,\varepsilon_1\rangle...|k_N,\varepsilon_N\rangle=\delta\left(\sum k_i\right)\;\prod_{i=1}^N\varepsilon^{(i)}_{\mu\nu} P_{\text{cl}}^{\mu}(z_i)P_{\text{cl}}^{\nu}(z_i),
$$
which is the result sought. This truncated description of the vertex can be seen to arise from (\ref{SFT}) in the special case where the states are of the form
$$
|\Phi\rangle=\varepsilon^{\mu\nu}\alpha_{-n\mu}\alpha_{-n\nu}|k\rangle
$$
where $n>0$. It is not too hard to show that, for such states the only non-vanishing contributions come from terms involving
$$
S_{n0}(z_i,z_j)=\frac{h'_j(0)}{(n-1)!}\left[\frac{\partial^{n-1}}{\partial t_i^{n-1}}\frac{1}{h_i(t_i)-z_j}\right]_{t_i=0},
$$
for $n>0$. When considering the scattering of such states, the vertex may be written in a truncated form
\begin{eqnarray}\label{21}
\langle\Sigma_N\hspace{-.1cm}:\hspace{-.1cm}X,P||\Phi_1\rangle...|\Phi_N\rangle&=&\int\prod_i\rd^Dp_i\;\delta\left(\sum p_i\right)\,\langle p_1|..\langle p_N|\; \exp\left(\sum_{i\neq j}S_{n0}(z_i,z_j)\,\alpha_0^{(j)}\cdot\tilde{\alpha}_n^{(i)}\right)\nonumber\\
&&\times\varepsilon_{(1)}^{\mu\nu}\,\alpha^{(1)}_{-n\mu}\,\alpha^{(1)}_{-n\nu}\,|k_1\rangle\,...\,\varepsilon_{(N)}^{\mu\nu}\,\alpha^{(N)}_{-n\mu}\,\alpha^{(N)}_{-n\nu}\,|k_N\rangle,
\end{eqnarray}
for fixed $n$, where all superfluous operators, i.e. those which annihilate on the physical states, are neglected. The only such state with the required conformal weight (recall $X(z)$ is of weight zero) is where $n=1$. Setting $n=1$ in (\ref{21}) and using the fact that $S_{10}(z_i,z_j)=(z_i-z_j)^{-1}$, the vertex (\ref{SFT}) reduces to the truncated vertex (\ref{X}) found above. The application of the analysis in Appendix A to the ambitwistor string also supports this result.

In summary, the $(X,P)$ contribution to the vertex is given by (\ref{SFT}). Upon restriction to physical external states one may use the truncated vertex (\ref{X}) without loss of generality. Momentum conservation, the result of integrating out the $X$ zero modes in the path integral approach, appears naturally and the requirement that $P(z_i)$ takes the value $P_{\text{cl}}(z_i)$ is ensured by the action of operators in the exponent in $\langle\Sigma_N|$ on the physical states. This completes the discussion of matter sector of the bosonic theory. The ghost sectors are treated in the following section. As a final comment, note that the gauge (\ref{gauge}) has been used throughout. It would be interesting to study the formalism in other gauges.

\subsection{The ghost sectors}

The conventional $(b,c)$ ghost system will be considered first. The $(\tilde{b},\tilde{c})$ system may be understood in a very similar way. Including the $b,c$ ghosts in the vertex modifies (\ref{Me}) to
$$
Z[J,K]\sim \int\rd^Dx\;e^{-S[X_{\text{cl}},P_{\text{cl}}]+S_0},
$$
where
\begin{equation}\label{SS}
S[X_{\text{cl}},P_{\text{cl}},b_{\text{cl}},c_{\text{cl}}]=-\int_{\Sigma}P_{\text{cl}}\cdot \bar{\partial}X_{\text{cl}}+b_{\text{cl}}\bar{\partial}c_{\text{cl}},	\quad	S_0=-\int_{\Sigma}x\cdot\bar{\partial}P_{\text{cl}}+\bar{\partial}b_{\text{cl}}(c_{-1}z^2+c_0z+c_1),
\end{equation}
and $S_0$ now contains zero modes for $X^{\mu}(z)$ and $c(z)$. Integrating out the $x^{\mu}$ zero mode gives the expected momentum conservation as seen above. The $c$ zero modes give contributions related to projective invariance and shall be discussed below. The operator contribution here is identical to that of the holomorphic sector of the conventional string. The ghost fields $b^{(i)}$ and $c^{(i)}$ have standard expansions when expressed in the local $t_i$ coordinates; however, their expansion is more complicated when expressed in the coordinate $z$ (see \cite{DiVecchia:1986mb} for further details)
$$
b^{(i)}(z)=(h_i'(z))^{-2}\sum_{n}b^{(i)}_nh^{-1}_i(z)^{-n-2}+...,	\qquad	c^{(i)}(z)=h_i'(z)\sum_{n}c^{(i)}_nh^{-1}_i(z)^{-n+1},
$$
and similarly for the $(\tilde{b},\tilde{c})$ ghosts. In order to compute the operator contribution to the amplitude it is useful to consider the Greens function $\left\langle c(z)b(w)\right\rangle=(z-w)^{-1}$, where the normalisation $\left\langle c_{-1}c_0c_1\right\rangle=1$ has been adopted for each set of ghosts. As in the conventional string theory, it is useful to introduce the following notation for the ghost vacuum, $\langle 3|=\langle 0|c_1c_0c_{-1}$. The contribution to the vertex from the $(b,c)$ system is then
\begin{equation}\label{bc}
\langle\Sigma_N\hspace{-.1cm}:\hspace{-.1cm}b,c|=\langle 3|_1\otimes...\otimes\langle 3|_N\;\exp\Bigg(\sum_{i,j}\sum_{\substack{n\geq 2\\ m\geq -1}}{\cal K}_{nm}(z_i,z_j)\,c_n^{(i)}\,b_m^{(j)}\Bigg)\prod_{n=-1}^{+1}{\cal Z}_n\prod_{I=1}^{N-3} b_I,
\end{equation}
where ${\cal K}_{nm}(z_i,z_j)$ is given in Appendix A. The ${\cal Z}_n$ are the same as given in (\ref{H}) and come from the zero mode contributions (\ref{SS}) discussed in the previous section. The required $N-3$ $b$ ghost insertions have also been included with the $b_I$ given by (\ref{b}). Integration over the moduli corresponding to the location of $N-3$ puncture then removes the $c$ ghosts from $N-3$ of the external states.

The story for the $(\tilde{b},\tilde{c})$ system is similar since they have the same conformal weights. The crucial difference is the gauge-fixing of the $e$ field leads to and extra $P^2$ term that must be taken into account as described in \cite{Adamo:2013tsa}. The $(\tilde{b},\tilde{c})$ contribution to the vertex is
\begin{equation}\label{tilde}
\langle\Sigma_N\hspace{-.1cm}:\hspace{-.1cm}\tilde{b},\tilde{c}|=\langle \widetilde{3}|_1\otimes...\otimes\langle  \widetilde{3}|_N\; \exp\Bigg(\sum_{i,j}\sum_{\substack{n\geq 2\\ m\geq -1}} {\cal K}_{nm}(z_i,z_j)\,\tilde{c}_n^{(i)}\,\tilde{b}_m^{(j)} \Bigg)\prod_{n=-1}^{+1}\widetilde{\cal Z}_n\prod_{I=1}^{N-3}\tilde{b}_I \; \bar{\delta}({\cal P}_I),
\end{equation}
where $\widetilde{\cal Z}_n$ is given by (\ref{H}), except with the $b$ ghost replaced by a $\tilde{b}$ ghost. In order to streamline the expressions the notation
$$
{\cal P}_I:=\int_{\Sigma}\mu_IP^2,
$$
has been introduced. A judicious choice of basis $\mu_I$ \cite{Adamo:2013tsa} yields ${\cal P}_I=\text{Res}_{z_I}(P^2)$ which gives the required $\bar{\delta}(k_I\cdot P(z_I))$ factor.

\subsection{Computing the amplitude}

It will be useful to denote the product of (\ref{X}), (\ref{bc}) and (\ref{tilde}) by\footnote{Note the ${\cal P}_I$ factors have been pulled out of $\langle\Sigma_N\hspace{-.1cm}:\hspace{-.1cm}\tilde{b},\tilde{c}|$ for later convenience.}
$$
\langle V_N|=\langle \Sigma_N| \;\prod_{I=1}^{N-3} b_I\;\tilde{b}_I \;\bar{\delta}({\cal P}_I),
$$
where
$$
\left\langle \Sigma_N\right|=\int\prod_{i=1}^N\rd p_i \;\delta\left(\sum p_i\right)\;\langle p,3,\tilde{3}|_1\otimes ...\otimes \langle p,3,\tilde{3}|_N\;\exp({\cal S})\prod_{n=-1}^{+1}{\cal Z}_n\prod_{m=-1}^{+1}\widetilde{\cal Z}_m.
$$
The exponent ${\cal S}$ is a sum of the oscillator exponent terms as given in (\ref{X}), (\ref{bc}) and (\ref{tilde})
$$
{\cal S}=\sum_{i,j} S(z_i,z_j)\,\alpha_0^{(j)}\cdot\tilde{\alpha}_1^{(i)} +\sum_{i,j} \sum_{\substack{n\geq 2\\ m\geq -1}} \left({\cal K}_{nm}(z_i,z_j)\,c_n^{(i)}\,b_m^{(j)} + {\cal K}_{nm}(z_i,z_j)\,\tilde{c}_n^{(i)}\,\tilde{b}_m^{(j)} \right).
$$
where, without loss of generality, the truncated vertex for the matter sector has been used. In terms of this vertex, the amplitude is given by
$$
{\cal M}(1,...,N)=g^{N-2}\int_{\Gamma\subset{\cal M}_{0,N}}\langle \Sigma_N| \;\prod_{I=1}^{N-3}b_I\,\tilde{b}_I \; \bar{\delta}\left({\cal P}_I\right) |\epsilon_1,k_1\rangle... |\epsilon_N,k_N\rangle,
$$
where the integral over the moduli space includes the (holomorphic) coordinates of $N-3$ punctures and where the physical vertex states are
$$
|\epsilon,k\rangle=c_1\tilde{c}_1\,\epsilon_{\mu\nu}\alpha_{-1}^{\mu}\alpha_{-1}^{\nu}\,e^{ik\cdot x}|0\rangle.
$$
In this expression $g$ is the coupling. The $b$ and $\tilde{b}$ insertions remove the $c$ and $\tilde{c}$ insertions on $N-3$ vertex operators and the $P^2$ insertion in ${\cal P}_I$ enforces the $\bar{\delta}(k\cdot P)$ at $N-3$ punctures, giving
$$
{\cal M}(1,...,N)=g^{N-2}\int_{\Gamma\subset{\cal M}_{0,N}} \langle \Sigma_N|\,\prod_{i=1}^3c_1^i\tilde{c}_1^i\epsilon^{\mu_i\nu_i}_i\alpha^{(i)}_{-1\mu_i}\alpha^{(i)}_{-1\nu_i}\,\prod_{i=4}^{N}\bar{\delta}\left(k\cdot P\right)\epsilon^{\mu_i\nu_i}_i\alpha^{(i)}_{-1\mu_i}\alpha_{-1\nu_i}^{(i)} |k_1\rangle...|k_N\rangle.
$$
In the previous sections it was shown that the $\alpha_{-1}^{(i)}$ factors in the external states are converted into $P_{\text{cl}}(z_i)$ factors by $\langle\Sigma_N|$. Momentum conservation is ensured by the contraction of the $\left| k_i\right\rangle$ with $\langle \Sigma_N|$. The only outstanding features of the bosonic theory to account for come from the ghost terms. The $\prod_{r=-1}^{+1}{\cal Z}_r$ combine with the $\prod_{i=1}^3c_1^i\tilde{c}_1^i$ terms, coming from three of the physical states, to give a contribution proportional to the inverse volume element of the holomorphic part of the conformal Killing group; SL($2;\C$), in line with what is found for the conventional string theory\footnote{
\begin{eqnarray}
\prod_{r=-1}^{+1}\sum_{m\geq 1}\sum_{i=1}^N{\cal M}_{rm}(z_i)b_m^i \;c_1^1c_1^2c_1^3|0\rangle_1...|0\rangle_N&=&\sum_{i<j<k}(z_i-z_j)(z_j-z_k)(z_k-z_i)\frac{b_{-1}^ib_{-1}^jb_{-1}^k}{h'_i(0)h'_j(0)h_k'(0)}  c_1^1c_1^2c_1^3|0\rangle_1...|0\rangle_N\nonumber\\
&=&\frac{(z_1-z_2)(z_2-z_3)(z_3-z_1)}{h'_1(0)h'_2(0)h_3'(0)} \left.|0\right\rangle_1...\left.|0\right\rangle_N\propto \frac{1}{\rd \text{Vol(SL($2;C$))}}\nonumber.
\end{eqnarray}
The last term is the inverse volume element of $SL(2;\C)$. Note that we have neglected the $h'_i(0)$ terms as such factors are guaranteed to cancel out in the final, conformally invariant, result.}. A similar calculation for the $\tilde{b}$ ghosts in $\widetilde{\cal Z}_n$ and the $\tilde{c}$ ghosts appearing in the physical states results in an additional inverse volume factor. The net result is the bosonic scattering amplitude \cite{Mason:2013sva}
$$
{\cal M}(1,...,N)=\delta^D\left(\sum k_i\right)\int_{\Gamma\subset{\cal M}_{0,N}} \frac{1}{\rd V_{123}}\prod_{i=1}^N\epsilon_i^{\mu\nu}P_{\mu}P_{\nu}\prod_i'\bar{\delta}\left(k_i\cdot P(z_i)\right),
$$
where
\begin{equation}\label{Vol}
\rd V_{ijk}= \frac{\rd z_i\rd z_j\rd z_k}{(z_i-z_j)(z_j-z_k)(z_k-z_i)},
\end{equation}
and
\begin{equation}\label{prod}
\prod_{i}' \bar{\delta}(k_i\cdot P(z_i))=\frac{1}{\rd V_{123}}\prod_{i=4}^N \bar{\delta}(k_i\cdot P(z_i)),
\end{equation}
is known to be permutation invariant \cite{Cachazo:2013hca}. Due to the action of the operators in the vertex, the momenta $P(z_i)$ above are all of the form (\ref{P}). As discussed in \cite{Mason:2013sva}, in order to construct amplitudes which includes Einstein gravity as a sub-sector requires the supersymmetric extension of the above bosonic construction. This is the subject of the next section.

\section{Incorporating fermions}

The bosonic sector of the ambitwistor theory is somewhat simpler than the corresponding bosonic sector of the conventional string theory. In many ways the fermionic sector plays a far more significant role in ambitwistor calculations. In this section it will be shown that the Neveu-Schwarz fermion contribution to the vertex operator is $\langle\Sigma_N\hspace{-.1cm}:\hspace{-.1cm}\Psi_1|\otimes \langle\Sigma_N\hspace{-.1cm}:\hspace{-.1cm}\Psi_2|$, where each $\langle\Sigma_N\hspace{-.1cm}:\hspace{-.1cm}\Psi|$ factor is given by
\begin{equation}\label{F}
\langle\Sigma_N\hspace{-.1cm}:\hspace{-.1cm}\Psi_r|=  \langle 0|_1...\langle 0|_N \exp\left(\sum_{i,j} \sum_{m,n\geq 1/2} S_{mn}(z_i,z_j)\,\psi_n^{(i)r}\cdot\psi_m^{(j)r}\right).
\end{equation}
This will be justified explicitly below. Since the fermionic components of the string and ambitwistor string worldsheet theories are comparable, one does not expect the fermion component of the vertex to differ in generalities. The main difference is that in the Type II ambitwistor string there are two copies of chiral fermions $\Psi^r$. In what follows only one copy will be explicitly considered and the $r=1,2$ index will be suppressed. It will be understood that two copies of the sector are needed for any explicit calculation. Only the Neveu-Schwarz sector is considered here. It is straightforward to  adapt existing constructions for the conventional superstring \cite{LeClair:1987de, AlvarezGaume:1988sj} to describe the Ramond sectors.

In the introduction the conserved charge approach of constructing the vertex was oultined. This approach is ideal for generalising to higher genus Riemann surfaces; however, a more direct approach may be employed for the genus zero case. For the free field constructions considered here, it is relatively straightforward to directly construct a vertex that will reproduce the correct scattering amplitudes by considering the two-point functions of the fields.

Following \cite{LeClair:1988sp, LeClair:1988sj, LeClair:1987de}, local holomorphic coordinates $t_i$ are introduced in a small disc around the $i$'th puncture. As usual, the puncture is taken to be at the origin of this coordinate system. In addition, there is the coordinate $z$ on the complex plane, related to the $t_i$ coordinates by the conformal transformation\footnote{For the three-point amplitude it is useful to use the SL$(2;\C)$ invariance to fix the three points at the canonical values $(0,1,\infty)$.} $z=h_i(t_i)$. The $i$'th puncture is then located at the point $z_i=h_i(0)$. Again, the idea is to map each of the individual local descriptions around the punctures to a common complex plane, where calculations are performed. The vertex function then takes the form
\begin{equation}\label{NS}
S_{mn}(z_i,z_j)=\oint_{t_i=0}\rd t_i\oint_{t_j=0}\rd t_j\, t_i^{-m-\frac{1}{2}}t_j^{-n-\frac{1}{2}}\;\sqrt{ h'_i h'_j}\;\frac{1}{h_i(t_i)-h_j(t_j)}.
\end{equation}
A more detailed discussion of the derivation of such vertex functions may be found in the Appendices. The vertex functions given in the previous section may be found in a similar way.

\subsection{Contribution to scattering amplitudes}

It will be shown in this section that this fermion vertex  reproduces the correct Pfaffian factor in the Neveu-Schwarz scattering amplitude. The starting point is the fermionic part of the vertex operator
$$
V(z)\sim\epsilon_{\mu}\Psi^{\mu}k_{\nu}\Psi^{\nu}\;e^{ik\cdot X(z)}.
$$
The state given by inserting the operator at the origin is $|k,\epsilon\rangle =\lim_{z\rightarrow 0}:V(z)| 0\rangle$ so that the states of interest contain terms of the form
$$
|k,\epsilon\rangle =\epsilon_{\mu}\psi_{-\frac{1}{2}}^{\mu}k_{\nu}\psi_{-\frac{1}{2}}^{\nu}| 0\rangle,
$$
where the expansion in terms of the oscillators $\psi_n^{\mu}$ is given with respect to a local coordinate system adapted to the puncture, and all dependence on other fields is temporarily supressed. The contribution from a single chiral fermion to the scattering amplitude of $N$ such states $|k_i,\epsilon_i\rangle$ is then
\begin{eqnarray}
&&{\cal M}_{\Psi}(1,2,...,N)=\langle\Sigma_N\hspace{-.1cm}:\hspace{-.1cm}\Psi||k_1,\epsilon_1\rangle...|k_N,\epsilon_N\rangle\nonumber\\
&=& \langle 0_1|...\langle 0_N|\exp\left(\sum_{i,j} \sum_{m,n\geq \frac{1}{2}} S_{mn}(z_i,z_j)\,\psi_m^{(i)}\cdot\psi_n^{(j)}\right) \prod_{r=1}^N\left(\epsilon_r\cdot\psi^{(r)}_{-\frac{1}{2}}k_r\cdot\psi^{(r)}_{-\frac{1}{2}}\right)   |0_1\rangle... | 0_N\rangle\nonumber.
\end{eqnarray}
For $m,n\neq\frac{1}{2}$ the $\psi_n^{(r)}$ oscillators anti-commute past the $\psi^{(r)}_{-\frac{1}{2}}$ states and annihilate on the vacuua $|0\rangle$, leaving
$$
{\cal M}_{\Psi}(1,2,...,N)= \left\langle 0_1 \right|...\left\langle 0_n \right|\exp\left(\sum_{i,j}  S_{\frac{1}{2}\frac{1}{2}}^{ij}\,\psi_{\frac{1}{2}}^i\cdot\psi_{\frac{1}{2}}^j\right) \prod_{r=1}^n\left(\epsilon_r\cdot\psi^{(r)}_{-\frac{1}{2}}\,k_r\cdot\psi^{(r)}_{-\frac{1}{2}}\right)    |\left. 0_1\right\rangle... |\left. 0_n\right\rangle.
$$
It is not hard to show that $S_{\frac{1}{2}\frac{1}{2}}(z_r,z_s)=\sqrt{h'_r(0)h'_s(0)}\left[h_r(0)-h_s(0)\right]^{-1}$. The factor of $\sqrt{h'_r(0)h'_s(0)}$ will cancel out in the final expression since the full amplitude is conformally invariant and so we shall not bother to keep track of such factors from now on.  Only the $m=n=1/2$ terms contribute so, in order to remove unnecessary clutter, it is helpful to denote $S_{\frac{1}{2}\frac{1}{2}}(z_i,z_j)$ and $\psi^{(i)}_{\frac{1}{2}}$ by $S(z_i,z_j)$ and $\psi^i$ respectively henceforth. Recalling that $h_i(0)=z_i$, the location of the $i$'th puncture, gives $S(z_i,z_j)=(z_i-z_j)^{-1}$. In order to evaluate the above operator expression it is useful to set
$$
\psi_{-m\mu}^{(i)}\rightarrow\eta_{\mu\nu}\frac{\partial}{\partial\psi_{m\nu}^{(i)}},	\qquad	m\geq\frac{1}{2}.
$$
The anti-commutation relations $\{\psi_{m\mu}^{(i)},\psi_{n\nu}^{(j)}\}=\delta_{ij}\eta_{\mu\nu}\delta_{m+n,0}$ are then satisfied automatically. The role of the vacua is then played by setting $\psi_m$ equal to zero once all derivatives have been evaluated. The contribution to the amplitude is then equivalent to
$$
{\cal M}_{\Psi}(1,...,N)=\left[ \prod_{i=1}^N \epsilon_{\mu}^i k_{\nu}^i 
\frac{\partial^2}{\partial \psi_{\mu}^i\partial\psi_{\nu}^i}\; 
\exp\left(\sum_{i,j} S(z_i,z_j)\,\psi^i\cdot\psi^j\right) \right]_{\psi=0}.
$$
Defining $\psi^{\mu}:=\epsilon^{\mu}\xi+k^{\mu}\tilde{\xi}$ for some grassmann variables $\xi $ and $\tilde{\xi}$ the differential operator may be written as
$$
\epsilon_{\mu} k_{\nu} 
\frac{\partial^2}{\partial \psi_{\mu}\partial\psi_{\nu}}=\frac{\partial^2}{\partial \xi\partial\tilde{\xi}},
$$
and the expression becomes
$$
{\cal M}_{\Psi}(1,...,N)=\left[ \prod_{i=1}^N \frac{\partial^2}{\partial \xi_i\partial\tilde{\xi}_i}\; 
\exp\left(\sum_{i,j} \left(
\begin{array}{cc}
\xi_i, & \tilde{\xi}_i
\end{array}\right) M'
\left(\begin{array}{c}
\xi_j \\
\tilde{\xi}_j
\end{array}
\right)
\right) \right]_{\xi=0},
$$
where the $2N\times 2N$ matrix $M$ is given by
$$
M'=\left(\begin{array}{cc}
A & C' \\
-C'^T & B
\end{array}\right),
$$
where,
$$
A=\frac{\epsilon_i\cdot\epsilon_j}{z_i-z_j},	\qquad	C'=\frac{\epsilon_i\cdot k_j}{z_i-z_j},	\qquad	B=\frac{k_i\cdot k_j}{z_i-z_j},
$$
for $i\neq j$ and $A_{ii}=B_{ii}=C'_{ii}=0$. This expression is identical to one obtained directly from the path integral (see Appendix B) and so ${\cal M}_{\Psi}(1,..,N)\propto\text{Pf}(M')$. Thus the NS fermion contribution to the (truncated) vertex is
$$
\langle\Sigma_N\hspace{-.1cm}:\hspace{-.1cm}\Psi|=  \langle 0|_1 ...\langle 0|_N \exp\Bigg(\sum_{i\neq j}\frac{ \psi^i\cdot\psi^j}{z_i-z_j}\Bigg).
$$
For completeness recall from  \cite{Mason:2013sva} that the inclusion of the $\epsilon\cdot P$ terms in the integrated vertex operators means that there is a non-zero contribution to the diagonal of the $C'$ sub-matrix and we should replace $C'$ with a matrix $C$, where the off-diagonal components are equal to those of $C'$ but has non-vanishing diagonal components $C_{ii}=\sum_{i\neq j}\frac{\epsilon_i\cdot k_j}{z_i-z_j}$. The inclusion of superconformal ghosts refines the above argument and the contribution to the vertex is in fact $\langle 0|_1\otimes\langle 0|_2\otimes \langle\Sigma_{N-2}\hspace{-.1cm}:\hspace{-.1cm}\Psi|$, rather than $\langle\Sigma_N\hspace{-.1cm}:\hspace{-.1cm}\Psi|$. The vertex with one chiral fermion has been described explicitly. A second copy is required in the Type II ambitwistor string.

\subsection{Superconformal ghosts}

This section shall be brief since the role of the superconformal ghosts in the ambitwisor string is very similar to that in conventional string theory, the operator formalism for which has been studied at length (see, for example, \cite{AlvarezGaume:1988sj}). Following standard procedures outlined in \cite{Friedan:1985ge} for the conventional superstring and in \cite{Adamo:2013tsa} for the ambitwistor string two vertex operators, for different pictures, are introduced. The fixed vertex operators take the form\footnote{These vertex operators look very similar to their counterparts in conventional superstring theory but it is important to stress that the origin of the ghosts is quite different. There are two copies of the $\gamma$ ghost coming from the fact that the ambitwistor string, unlike the Type II superstring, has two left-moving supersymmetries in contrast to the single chiral and anti-chiral supersymmetries of the closed superstring. By contrast, the $\tilde{c}$ ghost has no analogue in the conventional string.} $c\tilde{c}\;\delta(\gamma_1)\delta(\gamma_2)\;U$ where
$$
U=\epsilon_1\cdot\Psi_1\;\epsilon_2\cdot\Psi_2\;e^{ik\cdot X},
$$
and the integrated vertex operators are $c\tilde{c}\,V$ where
$$
V=e^{ik\cdot X}\prod_{r=1}^2(\epsilon_r\cdot P+k\cdot\Psi_r\; \epsilon_r\cdot \Psi_r).
$$
The corresponding states are given by $|\chi_i\rangle=\lim_{t_i\rightarrow 0}:U(t_i)|0\rangle$ and $|\Phi_i\rangle=\lim_{t_i\rightarrow 0}:V(t_i)|0\rangle$. The vertex contribution of the superghosts takes the standard form as that of the holomorphic sector of the closed superstring in the operator formalism, details of which may be found for example in \cite{LeClair:1987de,AlvarezGaume:1988sj}. The superghost contribution takes the form $\langle V_{\beta,\gamma}|=\langle V_1|\otimes\langle V_2|$ where
$$
\langle\Sigma_N\hspace{-.1cm}:\hspace{-.1cm}\beta^r,\gamma^r|=\langle \text{gh}|\;\exp\left(\sum_{i,j}\sum_{m,n}{\cal B}_{nm}(z_i,z_j)\,\gamma^{r(i)}_n\,\beta^{r(j)}_m\right),
$$
where $\langle \text{gh}|$ is an $N$-fold product of ghost vacua of appropriate picture number. The key point is that, in order for the amplitude to give a non-zero result, we require the insertion of picture-changing operators whose net effect is to convert two of the $c\tilde{c}V$ external states into $c\tilde{c}\;\delta(\gamma_1)\delta(\gamma_2)\;U$ states. This has two effects \cite{Mason:2013sva}; the first is that the two sets of $\delta(\gamma_1)\delta(\gamma_2)$ factors give rise to a factor of ${\cal G}':=(z_i-z_j)^{-1}\sqrt{\rd z_i\rd z_j}$ where $z_i$ and $z_j$ are the locations of the two $U$ states, the second is that only the $N-2$, $V$ states contribute to the matrix $M$ discussed in the previous section. Thus the Pfaffian of interest is $\text{Pfaff}(M^{ij}_{ij})$, which is the Pfaffian of $M$ with $i$'th and $j$'th rows and columns omitted. Further details may be found in \cite{Mason:2013sva}. Incorporating the net effects of ghosts and superghosts, the amplitude may be written as
$$
{\cal M}(1,...,N)= \int_{\Gamma\subset{\cal M}_{0,N}}\langle {\cal W}_N||\chi_1\rangle\otimes|\chi_2\rangle\otimes|\Phi_3\rangle\otimes...\otimes |\Phi_N\rangle,
$$
where the (truncated) vertex is given by
\begin{eqnarray}
\langle{\cal W}_N|&=&\int\prod_{i=1}^N\rd^{10}p_i\;\delta\left(\sum p_i\right) {\cal G}'\, \langle p_1|...\langle p_N|\;\prod_{i} \bar{\delta}'(k_i\cdot P_{\text{cl}}(z_i))\nonumber\\
&&\times\exp\Bigg(\sum_{i\neq j=3}^N\frac{ \psi^i\cdot\psi^j}{z_i-z_j}+\sum_{i\neq j}\frac{\alpha_0^{(j)}\cdot\tilde{\alpha}_1^{(i)}}{z_i-z_j}\Bigg),
\end{eqnarray}
where ${\cal G}'$ is defined above and $\bar{\delta}'(k\cdot P)$ is given by (\ref{prod}). The physical states are
$$
|\chi\rangle=\prod_{r=1,2}\epsilon_r\cdot \psi_r|k\rangle,	\qquad	|\Phi\rangle=\prod_{r=1,2}(\epsilon_r\cdot \alpha_{-1}+\epsilon_r\cdot\psi_r\;k\cdot\psi_r)|k\rangle,
$$
where $|k\rangle=e^{ik\cdot x}|0\rangle$. It is straightforward to show that the above vertex leads to the amplitude (\ref{CHY}). For more general computations, the description of the vertex with the full compliment of oscillators made explicit; (\ref{SFT}) and (\ref{F}), should be used in place of the truncated expressions in the above vertex.

\section{Discussion}

The operator formalism for conventional bosonic string theory has been reviewed from a number of complimentary perspectives and adapted to describe and operator formalism for the ambitwistor string. This is only the beginning of the story; there are a number of directions for future work.

\subsection{Loops}

The construction outlined in the Introduction can be generalised to higher genus Riemann surfaces\footnote{There is also an interesting connection with integrable systems \cite{Ishibashi:1986bd,AlvarezGaume:1987cg}.} \cite{Vafa:1987es,AlvarezGaume:1988bg,AlvarezGaume:1988sj}. The derivation of the vertex in the higher loop case takes a similar form in that we search for a single-valued function $f_n$ that is holomorphic on $\Sigma\backslash\{p\}$ and has a pole of order $n$ at the point $p$; however, the Weierstrass gap theorem requires that $n> g$ where $g$ is the genus of $\Sigma$. A way out is to construct a multi-valued function\footnote{The multi-valued function is simply a generalisation of (\ref{4}) where $\ln(z_i-z_j)$ is replaced by the two-point function $\ln( E(z_i,z_j))$ on the genus $g$ surface, where $E(z_i,z_j)$ is the Prime form.} with pole of order $n\leq g$ which is then made single-valued by the addition of a anti-holomorphic piece. For a scalar field, the function $f_n$ takes the form
$$
f_n=\int^z\left(\eta_n(z)-A_n(\text{Im}\Omega)^{-1}(\omega-\bar{\omega})\right),
$$
where $\Omega$ is the period matrix associated to $\Sigma$, $\omega$ is an abelian differential with expansion
$$
\omega(z)=\sum_{n=1}^{\infty}A_nz^{n-1}\rd z,
$$
and
$$
\eta_n(z)=\frac{1}{(n-1)!} \frac{\partial}{\partial z} \left[\frac{\partial^{n-1}}{\partial y^{n-1}}\ln E(z,y)\right]_{y=0},
$$
where $E(z,y)$ is the Prime form \cite{Fay}. Generalising to $N$ punctures, the functions $N_{nm}(z_i,z_j)$ may then be written in terms of these objects evaluated at the punctures. As noted above, this approach requires the introduction of anti-holomorphic data, which seems at odds with the general philosophy of ambitwistor string theory; however, the simplicity seen in the bosonic matter sector at tree level extends to higher genus and progress can be made in the one-loop case \cite{RR}. Much of the existing approach to string theory in the operator formalism is expected to also work for the ambitwistor string; in particular, one may adapt the arguments of \cite{Vafa:1987es,AlvarezGaume:1988bg,AlvarezGaume:1988sj} to describe the ghost and fermionic sectors of the ambitwistor string. 

\subsection{String field theory}

A triangulation of the moduli space of Riemann surfaces can be used deconstruct the moduli space into basic components which may form the geometric building blocks (Feynman rules) of the interactions of a string field theory. The basic components are cylinders and the restricted $N$-hedra described in \cite{Kugo:1989tk,strebel} and references therein. A crucial role is played by what are often referred to as the `missing regions' $\mathscr{D}_N$. These are regions of the moduli space of the $N$-punctured sphere ${\cal M}_{0,N}$ which are not produced by gluing together $M<N$ point vertices and so have to be added in by hand as higher point fundamental interactions. The requirement of adding in a new interaction term at each $N$ to give a complete single cover the moduli space famously requires that the string field action is non-polynomial \cite{Kugo:1989aa}. The action may be writen schematically as \cite{Kugo:1989aa,Kugo:1989tk}
$$
S=\frac{1}{2}\Phi {\cal Q} \Phi+\sum_{N=3}^\infty \frac{g^{N-2}}{N!}\Phi^N,
$$
where ${\cal Q}$ is the BRST operator, contains fundamental vertices $\Phi^N$ of arbitrarily high order. The vertices are naturally described in the operator formalism by contracting a vertex with string states $|\Phi\rangle$
$$
\Phi^N=\langle {\cal V}_N||\Phi_1\rangle...|\Phi_N\rangle.
$$
The vertex here is (again schematically) related to (\ref{Vbos}), the vertex $\langle \Sigma_N|$  considered in the Introduction, by
$$
\langle{\cal V}_N|\sim \int_{\mathscr{D}_N} \langle \Sigma_N|\prod_Ib_I.
$$
 Once the operator ${\cal Q}$ and a triangulation of moduli space are known, the vertex provides the crucial ingredient in the construction of the string field theory. As discussed in \cite{Mason:2013sva}, the physical interpretation of the bosonic sector of the ambitwistor string is unclear. In particular, the scattering amplitudes do not describe conventional Einstein gravity and it is uncertain what value should be placed on the construction of a bosonic ambitwistor string field theory; however, some progress has been made recently in understanding superstring field theory \cite{Sen:2015uaa} and such methods may be adapted to the ambitwsitor theory. A particularly intriguing possibility is that the string field theory in general curved backgrounds may be tractable. We shall report on these and related issues elsewhere \cite{RR}. 

\subsection{Background independence}

One of the great challenges facing string theory is how to translate the successes of the perturbative framework into a single coherent and background independent picture of the spacetime of string theory. Indeed, it was one of the great hopes of string field theory that a second quantised construction might provide such a framework. Using the operator formalism, some progress has been made in understanding the space of conformal field theories relevant to string backgrounds by constructing connections on such spaces \cite{Ranganathan:1993vj,Ranganathan:1992nb} and in demonstrating that the formal structure of bosonic string field theory is invariant under infinitesimal changes of background \cite{Sen:1993mh,Sen:1993kb}; however, with the advent of M-Theory it became clear that non-perturbative effects play a central role in the full quantum description.

There is considerable evidence that type II ambitwistor string theory correctly describes perturbative type II (complexified) supergravity. There is hope that, given we already have a background independent description of classical supergravity, a study of how tree level ambitwistor string field theory gives rise to classical Einstein supergravity may shed light on the problem of background independence in conventional string theory. A full understanding of string theory requires consideration of quantum effects and it is unlikely that any analysis of the kind proposed here will lead to a description of M-Theory; however, we may learn much from a diffeomorphism-invariant description of the classical theory. In many ways, this may be likened to the study of super Yang-Mills with sixteen supercharges in five dimensions; we suspect that the quantum description of this theory is given by a novel superconformal theory in six dimensions \cite{Douglas:2010iu, Lambert:2010iw} but there is still a sensible classical description of the five-dimensional theory. A gauge-invariant description of, even classical, Yang-Mills has value in determining the geometric structure of the theory. These ideas shall be explored further elsewhere.

\begin{appendices}

\section{Vertex functions}

\subsection{Conformal maps and vertex functions}

A standard mode expansion of a primary field of dimension $d$ is
$$
\phi(t)=\sum_n\phi_n\,t^{-n-d}
$$
Under a conformal transformation $t\rightarrow z=h(t)$, the primary field transforms as
\begin{equation}\label{1}
\phi(t)\rightarrow h[\phi(t)]=(h'(t))^d\;\phi(h(t)).
\end{equation}
where $h'$ is the derivative of $h$ with respect to $t$. Writing this new description of the field in terms of the `old' coordinates $t$, the mode expansion may be written as
\begin{equation}\label{2}
h[\phi(t)]=\sum_nh[\phi_n]\,t^{-n-d}.
\end{equation}
where the mode coefficients may be found in the standard way
$$
h[\phi_n]=\oint_{t=0}\frac{\rd t}{2\pi i}\,t^{n+d-1}\,h[\phi(t)]
$$
which may be written in terms of the transformed field $\phi(z)=\phi(h(t))$ using (\ref{2}) as
\begin{equation}\label{3}
h[\phi_n]=\oint_{t=0}\frac{\rd t}{2\pi i}\,t^{n+d-1}\,(h'(t))^d\,\phi(h(t))
\end{equation}
For example, the dimension one field $\partial X^{\mu}(z)$ gives
$$
h[\alpha^{\mu}_{-n}]=\oint_{t=0}\frac{\rd t}{2\pi i}\,t^{n+d-1}\,h'(t)\,\partial X^{\mu}(h(t)).
$$
It is a straightforward application of the commutation relations to show that, for $m,n>0$,
$$
{\cal N}_{mn}(z_i,z_j)=\frac{1}{2n}\langle 0|\exp\left(\sum_{k,l}\sum_{p,q>0}{\cal N}_{pq}(z_i,z_j)\,\alpha^{(k)}_p\cdot\alpha^{(l)}_q\right)\;\alpha^{(i)}_{-m}\cdot\alpha^{(j)}_{-n}|0\rangle
$$
Since only the contributions where $p$ and $q$ equal $-m$ or $-n$ and only the $i$'th and the $j$'th Fock spaces play a role, the above expression may be written compactly as $\langle V_2||\Phi_i\rangle|\Phi_j\rangle$ and is determined by the two-point function $\langle \partial X^{(i)}(z) \partial X^{(j)}(w)$ as described in \cite{LeClair:1988sp}. If we take $\langle \partial X(z) \partial X(w)=-\eta^{\mu\nu}(z-w)^{-2}$ then we find, for $m,n>0$,
$$
\langle h_i[\alpha_{-n}^{\mu(i)}]\,h_j[\alpha^{\nu(j)}_{-m}]\rangle=\frac{1}{n} \oint_{0}\frac{\rd t_i}{2\pi i}\,t^{-n}\,h'_i(t_i) \oint_{0}\frac{\rd t_j}{2\pi i}\,t^{-m}\,h'_j(t_j)\,\frac{-\eta^{\mu\nu}}{\left(h_i(t_i)-h_j(t_j)\right)^2}.
$$
Vertex functions for other contractions may be found in a similar way. Using the ghost contraction $\langle b(z)c(w)\rangle =(z-w)^{-1}$ and (\ref{3}) it is not hard to show that
$$
{\cal K}_{nm}(z_i,z_j)= -\oint\frac{\rd t_i}{2\pi i}\oint\frac{\rd t_j}{2\pi i}\; t_i^{-n+1}t_j^{-m-2} \;\left(h_i'(t_i)\right)^2 \left(h_j'(t_j)\right)^{-1}\;\frac{1}{h_i(t_i)-h_j(t_j)}
$$
and the contribution from the $c$ zero modes is found straightforwardly from
$$
\int_{\Sigma}\rd^2z\bar{\partial} b_{\text{cl}} (c_{-1}z^2+c_0z+c_1)=\sum_{i=1}^N\oint_{z_i}b_{\text{cl}}^{(i)} (c_{-1}z^2+c_0z+c_1)
$$
Using the standard expansion $b^{(i)}(z)=(h'_i(t_i))^{-2}\sum_nb_n^{(i)}t^{-n-2}_i$ and changing the integral to local $t_i$ coordinates gives
$$
\int_{\Sigma}\rd^2z\,\bar{\partial} b_{\text{cl}}(c_{-1}z^2+c_0z+c_1)=\sum_i\sum_n{\cal M}_{nm}(z_i)b_m^{(i)}{\cal C}^n
$$
where $n=-1,0,+1$, ${\cal C}=(c_{-1},c_0,c_1)$ and
$$
{\cal M}_{nm}(z_i)=\oint_{t_i=0} \frac{\rd t_i}{2\pi i}\;t_i^{-m-2}(h_i'(t))^{-1}(h_i(t))^{n+1}
$$
Similarly, for the fermions $\psi^{\mu}$ for which $\langle \psi^{\mu}(z)\psi^{\nu}(w)\rangle=(z-w)^{-1}$ and (\ref{3}) we have
$$
{\cal S}_{nm}(z_i,z_j)= -\oint\frac{\rd t_i}{2\pi i}\oint\frac{\rd t_j}{2\pi i}\; t_i^{-n+\frac{1}{2}}t_j^{-m-\frac{1}{2}} \;\sqrt{h_i'(t_i) h_j'(t_j)}\;\frac{1}{h_i(t_i)-h_j(t_j)}
$$

\subsection{Vertex functions from the classical action}

We summarise here the calculation of the vertex functions from the perspective of the classical action, as described in \cite{DiVecchia:1986mb}. Introducing the source
$$
J_{\mu}(z)=\sum_{i=1}^Nk^{(i)}_{\mu}\delta^2(z-z_i)
$$
the $N$-tachyon scattering may be computed from the extended action
$$
S=\int_{\Sigma}\rd^2z \partial X\cdot\bar{\partial}X+J\cdot X
$$
where the classical solution satisfies $\partial\bar{\partial}X_{\text{cl}}^{\mu}=J^{\mu}$. By using the standard trick of introducing the Green's function $G(z,w)$ and integrating by parts to give the expression for the classical action
$$
S_{\text{cl}}=\int_{\Sigma\times\Sigma}\rd^2z\rd^2 w\, G(z,w)\,J(z)\cdot J(w)
$$
We may then replace $J^{\mu}$ with $\partial\bar{\partial}X_{\text{cl}}^{\mu}$ to give an expression for the amplitude in terms of the Greens function and the classical solutions. Note that this procedure does not rely on the form of $J_{\mu}$ and so generalises beyond the simple case of Tachyon scattering. We take the Greens function to split into holomorphic and anti-holomorphic parts and concentrate on the former. Integrating by parts we find
$$
S_{\text{cl}}=\sum_{i,j}\oint_{z_i}\rd z\oint_{w_j} \rd w\,\ln(z-w)\,\partial_zX_{\text{cl}}(z) \cdot \partial_wX_{\text{cl}}(w)
$$
where the discs around the locations of the punctures have been excised and the contour integrals are about the boundaries of these discs. Introducing the local coordinates about each puncture $z=h_i(t_i)$ and $w=h_j(t_j)$ and using (\ref{1}) and (\ref{2}) we write
$$
\partial X(z)=(h'(t))^{-1}\sum_nh[\alpha_n]\,t^{-n-1}
$$
which then gives\footnote{Were we have written $h[\alpha_n]$ as simply $\alpha_n$.}
$$
S_{\text{cl}}=\sum_{i,j}\sum_{m,n}\oint_0\rd t_i\oint_0 \rd t_j\,t^{-n-1}_i\,t_j^{-m-1}\,\ln(h_i(t_i)-h_j(t_j))\,\alpha_n^{(i)} \cdot \alpha_m^{(j)}
$$
The vertex function is then read off as
\begin{equation}\label{4}
{\cal N}_{mn}(z_i,z_j)=\oint_0\rd t_i\oint_0 \rd t_j\,t^{-n-1}_i\,t_j^{-m-1}\,\ln(h_i(t_i)-h_j(t_j))
\end{equation}
It is not hard then to show that the vertex functions (often called Neumann functions in the older literature) for the $X$ sector of the bosonic string are (see for example \cite{LeClair:1988sp})
\begin{eqnarray}
{\cal N}_{00}(z_i,z_j)&=&\left\{
\begin{array}{cc}
\ln|h_i'(0)|, & i=j\\
\ln|z_i-z_j)|, & i\neq j.
\end{array}
\right.\nonumber\\
{\cal N}_{0m}(z_i,z_j)&=&\frac{1}{m}\oint\frac{\rd t}{2\pi i}\;t^{-m}h'_i(0)\frac{1}{h_i(t_i)-z_j}\nonumber\\
{\cal N}_{nm}(z_i,z_j)&=& \frac{1}{mn}\oint\frac{\rd t_i}{2\pi i}\oint\frac{\rd t_j}{2\pi i}\; t_i^{-n}t_j^{-m} \;\left(h_i'(t_i)\right)^2 \left(h_j'(t_j)\right)^{-1}\;\frac{1}{\left(h_i(t_i)-h_j(t_j)\right)^2}\nonumber
\end{eqnarray}
where $m,n>0$. The first follows directly from (\ref{4}). The second and third follow from (\ref{4}) by noting that the contour integral around the integrand with order $n+1$ pole can be written as the integral around the derivative of the integrand (with order $n$ pole). The $i=j$ case requires a more careful treatment of the Greens function than has been presented here.

\section{Fermion Path Integral}

In this appendix we show how the fermion path integral gives rise to the required Pffafian. The fermion contribution to the vertex operator 
$$
V(z)\sim\epsilon_{\mu}\Psi^{\mu}k_{\nu}\Psi^{\nu}\;e^{ik\cdot X(z)} 
$$
We can write the fermionic part of the vertex operator in exponential form by suing the standard trick if introducing dummy grassmann variables $\xi$ and $\tilde{\xi}$ as
$$
V(z)\sim\left[\frac{\partial^2}{\partial\xi\partial\tilde{\xi}}\;\exp\left(\xi \epsilon\cdot\Psi+\tilde{\xi}k\cdot\Psi+ik\cdot X\right)\right]_{\xi=\tilde{\xi}=0}.
$$
Neglecting the $e^{ik\cdot X}$ factor, the fermion contribution to the amplitude may be written as
$$
\left[\prod_{i=1}^n\frac{\partial^2}{\partial\xi_i\partial\tilde{\xi}_i}   \int{\cal D}\Psi\;\exp\left(-\frac{1}{2}\int_{\Sigma}\Psi\bar{\partial}\Psi\right)\prod_{i=1}^n\;\exp\left(\xi_i \epsilon\cdot\Psi+\tilde{\xi}_ik\cdot\Psi\right)\right]_{\xi_i=\tilde{\xi}_i=0}
$$
Defining the current
$$
J_{\mu}(z)=\sum_{i=1}^n\left(\xi^i \epsilon^i_{\mu}+\tilde{\xi}^ik^i_{\mu}\right)\delta^2(z-z_i)
$$
the path integral expression may be written as
\begin{eqnarray}
 \int{\cal D}\Psi\;\exp\left(-\frac{1}{2}\int_{\Sigma}\Psi\bar{\partial}\Psi-\int_{\Sigma} J\cdot \Psi\right)=\nonumber\\
 \int{\cal D}\eta\;\exp\left(-\frac{1}{2}\int_{\Sigma}\rd^2z\eta\bar{\partial}\eta+\frac{1}{2}\int_{\Sigma\times\Sigma}\rd^2z\rd^2w J_{\mu}(z)S(z,w)J^{\mu}(w)\right)\nonumber
\end{eqnarray}
where we have defined
$$
\eta^{\mu}(z)=\Psi^{\mu}(z)+\int_{\Sigma}\rd^2wS(z,w)J^{\mu}(w)
$$
where $S(z,w)$ is the Green's function for $\bar{\partial}$. Having shifted the integration from ${\cal D}\Psi$ to ${\cal D}\eta$ and now performing the Gaussian integral gives a simple $(\det{\bar{\partial}})^5$ factor which we shall ignore for now on. The remaining expression for the path integrals, upon substitution for $J_{\mu}$ gives
$$
\left[\prod_{i=1}^n\frac{\partial^2}{\partial\xi_i\partial\tilde{\xi}_i}\;\exp\left(\frac{1}{2}\sum_{i,j=1}^n\left(\xi \epsilon_{\mu}+\tilde{\xi}k_{\mu}\right)S(z_i,z_j)\left(\xi \epsilon^{\mu}+\tilde{\xi}k^{\mu}\right)\right)\right]_{\xi=\tilde{\xi}=0}
$$
which may be written as
$$
\left[\prod_{i=1}^n\frac{\partial^2}{\partial\xi_i\partial\tilde{\xi}_i}   \;\exp\left(\frac{1}{2}\sum_{i,j=1}^n\Xi_I^tM_{IJ}\Xi_J\right)\right]_{\xi=\tilde{\xi}=0}
$$
where $A^t$ denotes the transpose of the matrix $A$. The $2n$-dimensional Grassmann basis is given by $\Xi^t_I=(\xi_i,\tilde{\xi}_i)$ and the $2n\times 2n$ matrix $M_{IJ}$ is given by
$$
M_{IJ}=\left(\begin{array}{cc}
\frac{\epsilon_i\cdot\epsilon_j}{z_i-z_j} & \frac{\epsilon_i\cdot k_j}{z_i-z_j} \\
\frac{k_i\cdot\epsilon_j}{z_i-z_j} & \frac{k_i\cdot k_j}{z_i-z_j}
\end{array}
\right)
$$
where we have specified the Greens function appropriate to the genus zero worldsheet; $S(z,w)=(z-w)^{-1}$. We write the exponent as a two-form $\omega=\frac{1}{2}M_{IJ}\Xi^I\wedge \Xi^J$, where $I=1,2,...,2n$. Expanding out the exponent gives
$$
\left[\prod_{i=1}^n\frac{\partial^2}{\partial\xi_i\partial\tilde{\xi}_i}   \;e^{\omega}\right]_{\xi=\tilde{\xi}=0}=\left[\prod_{i=1}^n\frac{\partial^2}{\partial\xi_i\partial\tilde{\xi}_i}\;\sum_{k=1}^n\frac{\omega^k}{k!}\right]_{\xi=\tilde{\xi}=0}
$$
The derivatives select out the top form, to give
\begin{eqnarray}
\left[\prod_{i=1}^n\frac{\partial^2}{\partial\xi_i\partial\tilde{\xi}_i}   \;e^{\omega}\right]_{\xi=\tilde{\xi}=0}&=&\frac{1}{n!}\omega^n\nonumber\\
&=&\text{Pf}(\omega)\Xi^1\wedge\Xi^2\wedge...\wedge\Xi^{2n}
\end{eqnarray}
where we have used a standard definition of the Pfaffian in the last line. Thus we see that
$$
\left[\prod_{i=1}^n\frac{\partial^2}{\partial\xi_i\partial\tilde{\xi}_i}   \int{\cal D}\Psi\;\exp\left(-\frac{1}{2}\int_{\Sigma}\Psi\bar{\partial}\Psi\right)\prod_{i=1}^n\;\exp\left(\xi \epsilon\cdot\Psi+\tilde{\xi}k\cdot\Psi\right)\right]_{\xi=\tilde{\xi}=0}\propto\text{Pf}(M_{IJ})
$$
where $M_{IJ}$ is given above.

\end{appendices}

\end{document}